\documentstyle[12pt,prd,aps,epsf,axodraw]{revtex}
\begin{document}
\baselineskip=17pt
\draft
\title{Muon Pair Production by Electron-Photon Scatterings }
\author{H. Athar$^{1,2,}$\footnote{E-mail: athar@phys.cts.nthu.edu.tw},
 Guey-Lin Lin$^{2,}$\footnote{E-mail: glin@cc.nctu.edu.tw} and
 Jie-Jun Tseng$^{2,}$\footnote{E-mail: geny.py86g@nctu.edu.tw}}
\address{$^1$Physics Division, National Center for Theoretical Sciences,
 Hsinchu 300, Taiwan \\
  $^2$Institute of Physics, National Chiao Tung University,
 Hsinchu 300, Taiwan}
\date{\today}
\maketitle
\begin{abstract}
\tightenlines

The cross section for muon pair productions by electrons
scattering over photons, $\sigma_{MPP}$, is calculated analytically in the
 leading order.
It is pointed out that for the center-of-mass energy range,
$s \geq  5 \, m^{2}_{\mu}$,
the cross section, $\sigma_{MPP}$ is less than $1 \, \mu $b.
The differential
energy spectrum for either of the
resulting muons is given for the purpose of
high-energy neutrino astronomy. An implication of our result for
a recent suggestion concerning the high-energy cosmic neutrino generation
through this muon pair is discussed.
\end{abstract}


High-energy neutrino astronomy is now a rapidly
developing field \cite{Cline:1999ez}. This entails the identification
of possible sources for high-energy cosmic neutrinos. In particular, non
hadronic
processes taking place in cosmos remain an interesting possible source for
high-energy neutrinos.

In this context, Kusenko and Postma have recently
suggested that, for a class of topological defects that may
produce ultrahigh-energy photons at high red shift,
the dominance of muon pair production (MPP) in
$e^{-}\gamma \rightarrow e^{-}\mu^{+}\mu^{-}$
over triplet pair production (TPP) in
$e^{-}\gamma \rightarrow e^{-}e^{+}e^{-}$ for $s \geq  5 \, m^{2}_{\mu}$
enables the MPP process to be an efficient mechanism for generating
high-energy cosmic neutrinos
\cite{Kusenko:2001fk}.
The electrons in the initial state of the above
processes are considered as originating from the electromagnetic cascade
generated by the ultrahigh-energy photons scattering over the cosmic
microwave background photons present at high red shift.
Subsequently, assuming that this ultrahigh-energy photon source at high red
shift is responsible for the observed flux of ultrahigh-energy cosmic rays,
it was pointed out that the decays of muons in MPP give
rise to high-energy cosmic neutrino flux $E^{2}_{\nu}\phi_{\nu}$ typically
peaking at
$E_{\nu} \sim 10^{17}$ eV
\cite{Postma:2001na}.
Concerning the cross sections of the above processes, we note that,
although TPP process
has been thoroughly studied before\cite{Anguelov:1999ck}, there is still
no explicit estimate
for $\sigma_{MPP}$ for $s \geq  5 \, m^{2}_{\mu}$ to our knowledge.
In order to verify whether MPP indeed dominates over TPP for
$s \geq  5 \, m^{2}_{\mu}$,
we analytically calculate the
 $\sigma_{MPP}$ for $5 \, m^{2}_{\mu}\, \leq s \leq  20 \, m^{2}_{\mu}$.
In this {\em Rapid
Communication}, we present some
details of the calculation and compare our result with the value quoted
 in Refs.\cite{Kusenko:2001fk,Postma:2001na}.
 Our conclusion is that, within the above
$s$ range, the cross section $\sigma_{MPP}$ we have obtained
is at least three orders of magnitude smaller
than the one quoted and used in
 Refs.\cite{Kusenko:2001fk,Postma:2001na}. This implies that MPP
can not be the
dominating high-energy neutrino generating process as suggested in
 Refs.\cite{Kusenko:2001fk,Postma:2001na}. This conclusion is
based upon the value of the ratio
$R$ defined as $R\, \simeq \sigma_{MPP}/(\eta_{TPP} \sigma_{TPP})$
\cite{Kusenko:2001fk}, where
$\eta_{TPP}(s) \simeq 1.768 (s/m^{2}_{e})^{-0.75}$ is the inelasticity
for TPP
and $ \sigma_{TPP}\, \simeq \alpha^{3}m^{-2}_{e}\left[3.11\ln{(s/m^{2}_{e})}
-8.07\right]$, both for
$s\, \gg \, m^{2}_{e}$ \cite{Anguelov:1999ck}.
 The $\eta_{TPP}$ is basically the average fraction of the incident energy
 carried by the final state positron.
 The original estimate of
Refs. \cite{Kusenko:2001fk,Postma:2001na}
gives $R\, \simeq 10^2$, which favors MPP process
as the dominating cosmic high-energy
neutrino
generating process.
Namely the electron energy attenuation length due to TPP process
is much longer than the interaction length of MPP process.
However, since the correct value
for $\sigma_{MPP}$ within the energy range $5 \, m^{2}_{\mu}\,
\leq s \leq  20 \, m^{2}_{\mu}$ is three orders of magnitude smaller, the
ratio $R$ becomes
less than $1$. Therefore MPP is no longer an effective mechanism for
generating high-energy cosmic neutrinos at the high red shift.

\begin{figure}
\begin{center}
{
\unitlength=1.0 pt
\SetScale{1.0}
\SetWidth{0.7}      
\scriptsize    
{} \qquad\allowbreak
\begin{picture}(95,79)(0,0)
\Text(15.0,70.0)[r]{$e$}
\ArrowLine(16.0,70.0)(58.0,70.0)
\Text(80.0,70.0)[l]{$e$}
\ArrowLine(58.0,70.0)(79.0,70.0)
\Text(57.0,60.0)[r]{$\gamma$}
\DashLine(58.0,70.0)(58.0,50.0){3.0}
\Text(80.0,50.0)[l]{$\bar{\mu}$}
\ArrowLine(79.0,50.0)(58.0,50.0)
\Text(54.0,40.0)[r]{$\mu$}
\ArrowLine(58.0,50.0)(58.0,30.0)
\Text(15.0,30.0)[r]{$\gamma$}
\DashLine(16.0,30.0)(58.0,30.0){3.0}
\Text(80.0,30.0)[l]{$\mu$}
\ArrowLine(58.0,30.0)(79.0,30.0)
\Text(47,0)[b] {diagram (a)}
\end{picture} \
{} \qquad\allowbreak
\begin{picture}(95,79)(0,0)
\Text(15.0,70.0)[r]{$e$}
\ArrowLine(16.0,70.0)(58.0,70.0)
\Text(80.0,70.0)[l]{$e$}
\ArrowLine(58.0,70.0)(79.0,70.0)
\Text(57.0,60.0)[r]{$\gamma$}
\DashLine(58.0,70.0)(58.0,50.0){3.0}
\Text(80.0,50.0)[l]{$\mu$}
\ArrowLine(58.0,50.0)(79.0,50.0)
\Text(54.0,40.0)[r]{$\mu$}
\ArrowLine(58.0,30.0)(58.0,50.0)
\Text(15.0,30.0)[r]{$\gamma$}
\DashLine(16.0,30.0)(58.0,30.0){3.0}
\Text(80.0,30.0)[l]{$\bar{\mu}$}
\ArrowLine(79.0,30.0)(58.0,30.0)
\Text(47,0)[b] {diagram (b)}
\end{picture} \
{} \qquad\allowbreak
\linebreak
\begin{picture}(95,79)(0,0)
\Text(15.0,70.0)[r]{$e$}
\ArrowLine(16.0,70.0)(37.0,60.0)
\Text(15.0,50.0)[r]{$\gamma$}
\DashLine(16.0,50.0)(37.0,60.0){3.0}
\Text(47.0,64.0)[b]{$e$}
\ArrowLine(37.0,60.0)(58.0,60.0)
\Text(80.0,70.0)[l]{$e$}
\ArrowLine(58.0,60.0)(79.0,70.0)
\Text(57.0,50.0)[r]{$\gamma$}
\DashLine(58.0,60.0)(58.0,40.0){3.0}
\Text(80.0,50.0)[l]{$\mu$}
\ArrowLine(58.0,40.0)(79.0,50.0)
\Text(80.0,30.0)[l]{$\bar{\mu}$}
\ArrowLine(79.0,30.0)(58.0,40.0)
\Text(47,0)[b] {diagram (c)}
\end{picture} \
{} \qquad\allowbreak
\begin{picture}(95,79)(0,0)
\Text(15.0,60.0)[r]{$e$}
\ArrowLine(16.0,60.0)(37.0,60.0)
\Text(47.0,61.0)[b]{$\gamma$}
\DashLine(37.0,60.0)(58.0,60.0){3.0}
\Text(80.0,70.0)[l]{$\mu$}
\ArrowLine(58.0,60.0)(79.0,70.0)
\Text(80.0,50.0)[l]{$\bar{\mu}$}
\ArrowLine(79.0,50.0)(58.0,60.0)
\Text(33.0,50.0)[r]{$e$}
\ArrowLine(37.0,60.0)(37.0,40.0)
\Text(15.0,40.0)[r]{$\gamma$}
\DashLine(16.0,40.0)(37.0,40.0){3.0}
\Line(37.0,40.0)(58.0,40.0)
\Text(80.0,30.0)[l]{$e$}
\ArrowLine(58.0,40.0)(79.0,30.0)
\Text(47,0)[b] {diagram (d)}
\end{picture} \
{} \qquad\allowbreak
\begin{picture}(95,79)(0,0)
\Text(15.0,70.0)[r]{$e$}
\ArrowLine(16.0,70.0)(37.0,60.0)
\Text(15.0,50.0)[r]{$\gamma$}
\DashLine(16.0,50.0)(37.0,60.0){3.0}
\Text(47.0,64.0)[b]{$e$}
\ArrowLine(37.0,60.0)(58.0,60.0)
\Text(80.0,70.0)[l]{$e$}
\ArrowLine(58.0,60.0)(79.0,70.0)
\Text(57.0,50.0)[r]{$Z$}
\DashLine(58.0,60.0)(58.0,40.0){3.0}
\Text(80.0,50.0)[l]{$\mu$}
\ArrowLine(58.0,40.0)(79.0,50.0)
\Text(80.0,30.0)[l]{$\bar{\mu}$}
\ArrowLine(79.0,30.0)(58.0,40.0)
\Text(47,0)[b] {diagram (e)}
\end{picture} \
{} \qquad\allowbreak
\begin{picture}(95,79)(0,0)
\Text(15.0,60.0)[r]{$e$}
\ArrowLine(16.0,60.0)(37.0,60.0)
\Text(47.0,61.0)[b]{$Z$}
\DashLine(37.0,60.0)(58.0,60.0){3.0}
\Text(80.0,70.0)[l]{$\mu$}
\ArrowLine(58.0,60.0)(79.0,70.0)
\Text(80.0,50.0)[l]{$\bar{\mu}$}
\ArrowLine(79.0,50.0)(58.0,60.0)
\Text(33.0,50.0)[r]{$e$}
\ArrowLine(37.0,60.0)(37.0,40.0)
\Text(15.0,40.0)[r]{$\gamma$}
\DashLine(16.0,40.0)(37.0,40.0){3.0}
\Line(37.0,40.0)(58.0,40.0)
\Text(80.0,30.0)[l]{$e$}
\ArrowLine(58.0,40.0)(79.0,30.0)
\Text(47,0)[b] {diagram (f)}
\end{picture} \
{} \qquad\allowbreak
\begin{picture}(95,79)(0,0)
\Text(15.0,70.0)[r]{$e$}
\ArrowLine(16.0,70.0)(58.0,70.0)
\Text(80.0,70.0)[l]{$e$}
\ArrowLine(58.0,70.0)(79.0,70.0)
\Text(57.0,60.0)[r]{$Z$}
\DashLine(58.0,70.0)(58.0,50.0){3.0}
\Text(80.0,50.0)[l]{$\mu$}
\ArrowLine(58.0,50.0)(79.0,50.0)
\Text(54.0,40.0)[r]{$\mu$}
\ArrowLine(58.0,30.0)(58.0,50.0)
\Text(15.0,30.0)[r]{$\gamma$}
\DashLine(16.0,30.0)(58.0,30.0){3.0}
\Text(80.0,30.0)[l]{$\bar{\mu}$}
\ArrowLine(79.0,30.0)(58.0,30.0)
\Text(47,0)[b] {diagram (g)}
\end{picture} \
{} \qquad\allowbreak
\begin{picture}(95,79)(0,0)
\Text(15.0,70.0)[r]{$e$}
\ArrowLine(16.0,70.0)(58.0,70.0)
\Text(80.0,70.0)[l]{$e$}
\ArrowLine(58.0,70.0)(79.0,70.0)
\Text(57.0,60.0)[r]{$Z$}
\DashLine(58.0,70.0)(58.0,50.0){3.0}
\Text(80.0,50.0)[l]{$\bar{\mu}$}
\ArrowLine(79.0,50.0)(58.0,50.0)
\Text(54.0,40.0)[r]{$\mu$}
\ArrowLine(58.0,50.0)(58.0,30.0)
\Text(15.0,30.0)[r]{$\gamma$}
\DashLine(16.0,30.0)(58.0,30.0){3.0}
\Text(80.0,30.0)[l]{$\mu$}
\ArrowLine(58.0,30.0)(79.0,30.0)
\Text(47,0)[b] {diagram (h)}
\end{picture} \
}
\end{center}
\caption{The Feynman diagrams contributing to MPP in the leading order.
\label{fig:feyman}}
\end{figure}
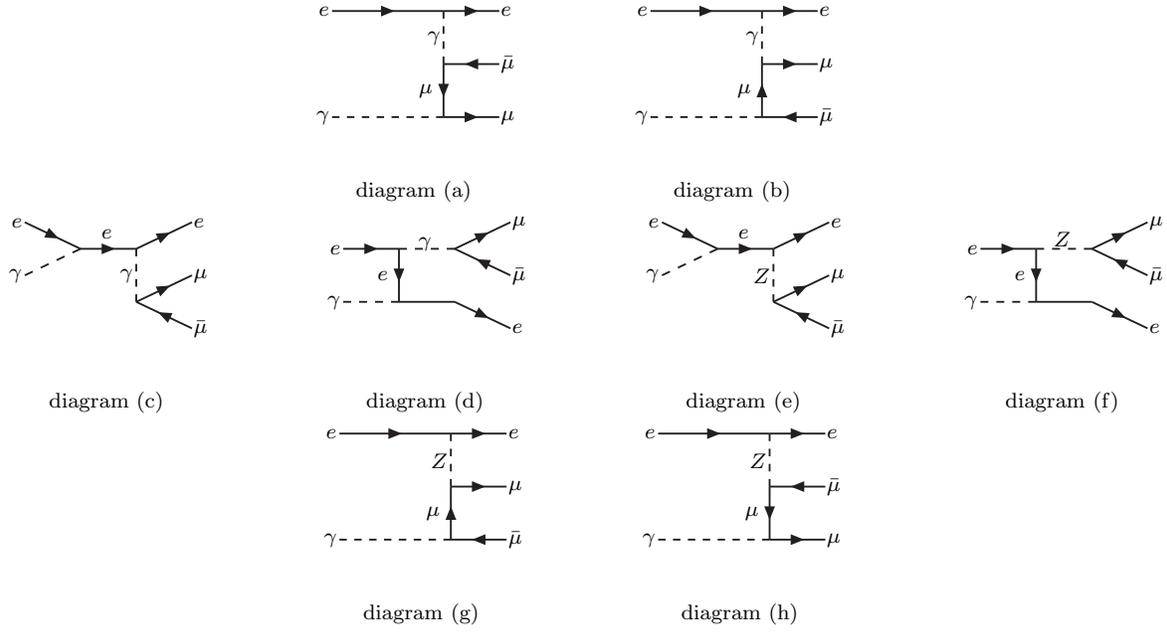

There are 8 Feynman diagrams contributing to the MPP process in the leading
order.
These are shown in Fig. \ref{fig:feyman}.
Among these 8 diagrams, 4 diagrams contain a $Z$-boson exchange, which
can be disregarded for the range of $s$ under
discussion. Among the remaining 4 diagrams, one can also disregard
diagrams (c) and (d) due to the inflow of large
invariant energies into the electron [diagram (c)] or
photon [diagram (d)] propagators.
We therefore only perform analytic calculations for
diagrams (a) and (b) in Fig. \ref{fig:feyman} while evaluate the rest
 of the diagrams numerically.
We have applied the equivalent photon
approximation \cite{Brodsky:1971ud} to
compute diagrams (a) and (b). We write the MPP cross section in the
following convolution
\begin{equation}\label{def}
\sigma_{MPP}=\int_{4m_{\mu}^2/s}^{1} dx f_{\gamma/e}(x)\sigma_
 {\gamma\gamma\to \mu^+\mu^-}(\hat{s}=xs),
\end{equation}
where $f_{\gamma/e}(x)=(\alpha/2\pi)[(1+(1-x)^2)/x]\ln(s/m^{2}_{e})$ is the
probability
of finding a photon from the incident electron with an energy fraction
$x=E_{\gamma}/E_e$.
The differential cross section as a function of the outgoing muon energy
$E_{\mu}$, in head on collisions, is given by
\begin{eqnarray}\label{final}
 \frac{\mbox{d}\sigma_{MPP}}{\mbox{d}y}\, &\simeq&
       \frac{\alpha^{3}}{m_{\mu}^2y}
\left[1+\left(1-\frac{4m_{\mu}^2y^2}{s}\right)^2\right]
 \ln{\left(\frac{s}{m^{2}_{e}}
 \right)}\nonumber \\
&\times&(1-v^2)\left[\left(1+\frac{1}{y^2}\right)
 \ln{\left(\frac{1+v}{1-v}\right)}
-v-\frac{1}{y^4}\left(\frac{v}{1-v}+\tanh^{-1}(v)\right)\right],
\end{eqnarray}
where $y=E_{\mu}/m_{\mu}$ with $y_{min}=1$ and $y_{max}=\sqrt{s}/2m_{\mu}$;
$v$ is the velocity of
the outgoing muon, which is related to $y$ by $v=\sqrt{1-\frac{1}{y^{2}}}$.
Apart from the trivial prefactor,
the first line in Eq.~(\ref{final}) arises from the
distribution function $f_{\gamma/e}(x)$ while the second line
describes the cross section for $\gamma\gamma\to \mu^+\mu^-$.
The total cross section $\sigma_{MPP}$, obtained by performing the $y$
integration, depends only on the center-of-mass energy $\sqrt{s}$.
For $s \, = \, 5\,
m_{\mu}^2$,
$\sigma_{MPP}\, \simeq 4\cdot 10^{-3}\, \mu$b,
while $\sigma_{MPP}$ increases to about
 $0.1 \, \mu$b for
 $s\,=\, 20 \, m_{\mu}^2$. These cross-section values are
at least 3 orders of magnitudes smaller than those of Refs.
 \cite{Kusenko:2001fk,Postma:2001na},
which give $\sigma_{MPP}$ between $0.1$ and $1$ mb in the above energy range.
The last equation gives the differential energy spectrum of either of the
muons produced in MPP. In the large $s$ limit, the $\sigma_{MPP}$
given by Eq.
(\ref{final})
behaves as
\begin{equation}\label{mpplarges}
 \sigma_{MPP}(s)\, \simeq \,  \frac{2\alpha^{3}}{m^{2}_{\mu}}\ln(2)
  \ln\left(\frac{s}{m^{2}_{e}}\right).
\end{equation}
The inelasticity for MPP (in the center-of-mass frame) is defined as
\begin{equation}\label{inelasticity}
 \eta_{MPP}(s)=\frac{1}{\sigma_{MPP}(s)}\int dE_{\mu}
 \left(\frac{E_{\mu}}{\sqrt{s}}\right)
 \frac{d\sigma_{MPP}}{dE_{\mu}}.
\end{equation}
Using Eqs. (\ref{final}) and (\ref{inelasticity}), we
find that the average fraction
of incident
energy carried by either of the muons in MPP, in the large $s$
limit, behaves as
\begin{equation}\label{larges}
 \eta_{MPP}(s) \, \simeq 3.44 \left(\frac{s}{m^{2}_{\mu}}\right)^{-0.5},
\end{equation}
whereas it is close to 0.48 near the threshold for the MPP process and
 approaches to  0.34 for  $s\, \sim 20 \, m^{2}_{\mu}$.

\begin{figure}[t]
\begin{center}
\leavevmode
\epsfxsize=3.5in
\epsfysize=3.5in
\epsfbox{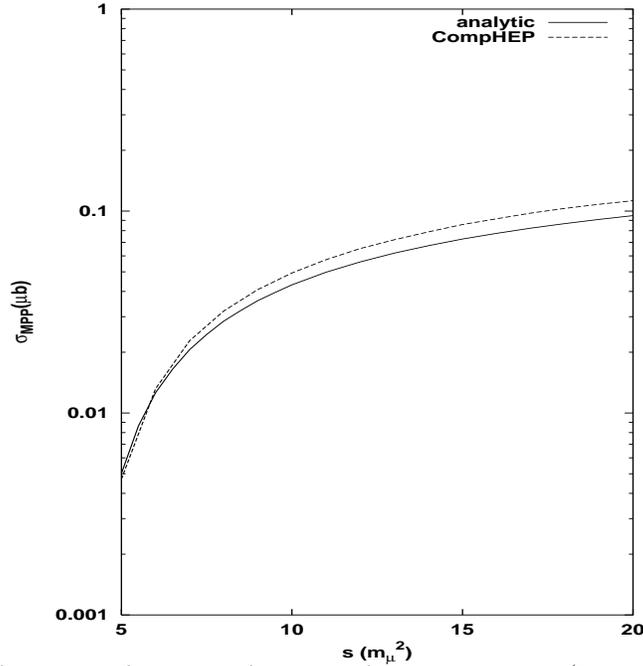}
\tightenlines
\caption{$\sigma_{MPP}$ as a function of square of center-of-mass
 energy  $s$
 (in units of $m^{2}_{\mu}$). Solid curve is obtained using Eq.
 (\ref{final}). Dashed curve is obtained using CompHEP (it
 includes contribution from all 8 diagrams, see [6]).
\label{fig:comparison}}
\end{center}
\end{figure}

We have used a package for
computations in high energy physics, CompHEP (version 33.22),
to check our results for $\sigma_{MPP}$
\cite{Pukhov:1999gg}. The comparison for
 $5 \, m^{2}_{\mu}\, \leq s \leq  20 \, m^{2}_{\mu}$ is shown in
Fig. \ref{fig:comparison}. We note that the plot generated by CompHEP is
a result of computing all 8 diagrams.
From Eq. (\ref{final}), we also note that, for
$s\, \gg \, 5 \, m^{2}_{\mu}$, the MPP cross section $\sigma_{MPP} <\, 1
 \, \mu $b.
We have confirmed
this remark using CompHEP as well.
We have made a further check on our
$\sigma_{MPP}$ by
 generating  the amplitude of diagrams (a) and (b) symbolically
using the package $FeynArts \, 3$\cite{Hahn:2000kx}. In this
procedure, we do not use the equivalent photon approximation. We
then obtained $\sigma_{MPP}$ by numerically performing the phase
space integration. We have found a rather good agreement (within
few percent) between the $\sigma_{MPP}$ obtained in this way and
that given by the equivalent photon approximation. Because of the
rather lengthy algebraic expressions occurring in the above
procedure, we are omitting further details of this check.

To obtain a consistent estimate of $R$, it is
desirable to compute $\sigma_{TPP}$ and $\eta_{TPP}$
in addition to
$\sigma_{MPP}$.
This can be easily done in the large $s$ limit.
By replacing $\mu^{\pm}$ with $e^{\pm}$ in the final state,
the same 8 diagrams shown in Fig. \ref{fig:feyman} also contribute to
the
TPP process in the leading order.
The dominating diagrams are again
the first two. In the equivalent photon
approximation with $s\gg m_e^2$, they give
\begin{equation}\label{tpp}
 \sigma_{TPP}(s)\, \simeq \, \frac{\alpha^{3}}{m^{2}_{e}}
 \ln(2)\ln\left(\frac{s}{m^{2}_{e}}\right),
\end{equation}
whereas
\begin{equation}\label{tpplarges}
 \eta_{TPP}(s) \, \simeq  3.44\left(\frac{s}{m^{2}_{e}}\right)^{-0.5}.
\end{equation}
The result for $\sigma_{TPP}$ can be easily inferred from
Eq. (\ref{mpplarges}) by replacing
$m_{\mu}^2$ there with $m_{e}^2$ and multiplying a
symmetry factor $1/2$ which takes into account the identical-particle
effect in the TPP process.
The above results for
$\sigma_{TPP}$ and $\eta_{TPP}$
agree well (within few percent) with the results quoted in
\cite{Anguelov:1999ck}.
We further evaluated $\eta_{TPP}(s)$ in the Lab
frame also,
which in the large $s$ limit, for the two dominating diagrams, is
$\sim 0.25(s/m^{2}_{e})^{-0.5}$.
In practice, one should use the Lab-frame $\eta_{TPP}$ in
the the definition of $R$.
Numerically, the Lab-frame
$\eta_{TPP}$ is approximately an order of magnitude smaller than its
center-of-frame counterpart. This suppression is
related to the kinematical factors present in the
Lorentz boost from the center-of-mass frame to the Lab-frame.
Using the Lab-frame $\eta_{TPP}$ and the values for $\sigma_{TPP}$ and
$\sigma_{MPP}$, we still
obtain $R\, < \, 1$ for
$5 \, m^{2}_{\mu}\, \leq s \leq  20 \, m^{2}_{\mu}$.

In summary, the authors in Refs. \cite{Kusenko:2001fk,Postma:2001na}
use the value 0.1-1 mb for the cross section $\sigma_{MPP}$ in the
energy range $5 \, m^{2}_{\mu}\, \leq s \leq  20 \, m^{2}_{\mu}$.
 As a result, they have deduced that
 $R\, \simeq \sigma_{MPP}/(\eta_{TPP} \sigma_{TPP})\, \gg 1$.
However, as we have shown, the
correct value for $\sigma_{MPP}$ obtained from Eq. (\ref{final}) yields
$R\, < 1$. In particular, near the threshold for the MPP process, i.e., for
$s\, \sim 5 \, m^{2}_{\mu}$, we have $R\, \ll \, 1$.
Therefore, based on this
observation, we conclude that MPP can not be a dominating high-energy
cosmic neutrino generating process as suggested in
 Refs.\cite{Kusenko:2001fk,Postma:2001na}. Furthermore, we have also derived
analytic expressions
(in the leading order) for differential energy spectrum and
the inelasticity of the MPP process in the large $s$ limit, which
might be
of relevance in some other contexts of high-energy neutrino astronomy.

    For completeness, let us add a final remark concerning
the possibility of high-energy cosmic neutrino generation in an
electromagnetic cascade at high red shift, $z$, through $\gamma
\gamma $ collisions. The process $\gamma \gamma \rightarrow
\mu^{+}\mu^{-}$ can, in principle, generate high-energy cosmic
neutrinos near the threshold for this process, namely, for
$\sqrt{s}\simeq 2 \, m_{\mu}$. The neutrinos are generated through
the subsequent decays of the final-state muons for a rather small
range of $z$ values, typically, for $z\sim 5-10$. In this $z$
range, the interaction length for the process $\gamma \gamma
\rightarrow \mu^{+}\mu^{-}$ is smaller than the energy attenuation
length dictated by the process $\gamma \gamma \rightarrow
e^{+}e^{-}$. Furthermore, this interaction length is also smaller
than the horizon length, $cH(z)^{-1}$, where $H(z)$ is the Hubble
constant in this $z$ range.

\acknowledgments
HA thanks Physics Division of NCTS for financial support.
GLL and JJT are supported by the National Science Council of
R.O.C. under the
grant number NSC89-2112-M009-041.

\pagebreak

%
%

%
%

\end{document}